\documentclass{emulateapj}
\usepackage{apjfonts}

\newcommand{\ddt}[1]{\frac{d #1}{d t}}
\newcommand{\sh}{\Sigma_{\mathrm{H}}}
\newcommand{\she}{\Sigma_{\mathrm{He}}}
\newcommand{\estarh}{E^{*}_{\mathrm{H}}}
\newcommand{\estarhe}{E^{*}_{\mathrm{He}}}
\newcommand{\estarb}{E^{*}_{\mathrm{rp}}}
\newcommand{\epsh}{\epsilon_{\mathrm{H}}}
\newcommand{\epsb}{\epsilon_{\mathrm{rp}}}
\newcommand{\epshe}{\epsilon_{\mathrm{He}}}

\newcommand{\lacc}{l_{\mathrm{acc}}}

\newcommand{\frp}{f_{\mathrm{rp}}}

\begin{document}

\title{On the Physics of Type I X-ray Bursts on Accreting Neutron
Stars at High Accretion Rates}
\shorttitle{Physics of X-ray Bursts}
\author{Randall L.\ Cooper and Ramesh Narayan}
\affil{Harvard-Smithsonian Center for Astrophysics, 60 Garden Street,
Cambridge, MA 02138}

\email{rcooper@cfa.harvard.edu, rnarayan@cfa.harvard.edu}

\begin{abstract}

We investigate the effect of the hot CNO cycle breakout reaction
$^{15}$O($\alpha$,$\gamma$)$^{19}$Ne on the occurrence of type I X-ray
bursts on accreting neutron stars.  For $\frp \lesssim 0.1$, where
$\frp$ is a dimensionless factor by which we multiply the
$^{15}$O($\alpha$,$\gamma$)$^{19}$Ne reaction rate of \citet{CF88},
our model predicts that bursts should occur only for accretion rates
$\dot{M}$ below a critical value $\approx 0.3 \dot{M}_{\mathrm{Edd}}$.
This agrees with observations.  For larger values of $\frp$, including
the standard choice $\frp = 1$, the model switches to a new regime in
which bursts occur all the way up to $\dot{M} \approx
\dot{M}_{\mathrm{Edd}}$.  Since the latter regime disagrees with
observations, we suggest that the true
$^{15}$O($\alpha$,$\gamma$)$^{19}$Ne reaction rate is lower than
usually assumed.

\end{abstract} 


\section{Introduction}

Type I X-ray bursts are thermonuclear explosions that occur on
accreting neutron stars in low-mass X-ray binaries.
They are triggered by thermally unstable H or He burning
near the stellar surface \citep[for reviews, see][]{C04,SB03}.  For
systems with accretion rates $\dot{M} \lesssim 0.1
\dot{M}_{\mathrm{Edd}}$, where $\dot{M}_{\mathrm{Edd}}$ denotes the
mass accretion rate at which the accretion luminosity is equal to the
Eddington limit, the basic physics of the burst onset is well
understood to be that of the thin shell thermal instability
\citep{SH65,HvH75}, and theoretical models have been rather successful
at reproducing the gross characteristics of burst observations 
in this regime \citep[e.g.,][]{FHM81,FL87,CB00,NH03}.  

This {\em Letter} addresses a longstanding problem that afflicts
nearly all burst models when the accretion rate $\dot{M} \gtrsim 0.1
\dot{M}_{\mathrm{Edd}}$.  Both simple one-zone burst models
\citep{FHM81,P83,B98,HCW05} and sophisticated time-dependent
multi-zone models \citep{AJ82,TWL96,FHLT03,HCW05} predict that bursts
should occur for all $\dot{M}$ up to $\approx \dot{M}_{\mathrm{Edd}}$.
Observations, however, indicate that bursts do not occur for $\dot{M}
\gtrsim 0.3 \dot{M}_{\mathrm{Edd}}$
\citep{vPCLJ79,vPPL88,Cetal03,RLCN06}.  Furthermore, \citet{vPPL88}
found that, for $0.1 \lesssim \dot{M}/\dot{M}_{\mathrm{Edd}} \lesssim
0.3$, a significant fraction of the accreted plasma burns stably
between consecutive bursts, leading to large values $\gtrsim 1000$ of
the parameter $\alpha$, the accretion energy released between
successive bursts divided by the nuclear energy released during a
burst.  Most theoretical models, on the other hand, predict that
nearly all of the accreted matter burns unstably during bursts, giving
a nearly constant $\alpha < 100$ at all $\dot{M}$.

\citet[][hereafter NH03]{NH03} developed a global linear stability
analysis of the accreted plasma on the surface of a neutron star.
They discovered a new regime of unstable nuclear burning for $0.1
\lesssim \dot{M}/\dot{M}_{\mathrm{Edd}} \lesssim 0.3$ that they
referred to as ``delayed mixed bursts.''  Their model reproduced both
the occurrence of considerable stable burning preceding a burst
(leading to values of $\alpha \gtrsim 1000$) and the absence of bursts
for $\dot{M} \gtrsim 0.3 \dot{M}_{\mathrm{Edd}}$, in agreement with
observations.  However, the complexity of their model made it
difficult to understand the basic physics behind delayed mixed bursts
and to identify the reasons why their results differed from those of
other theoretical models at high accretion rates.  To remedy this,
\citet[][hereafter CN06]{CN06} constructed a simple two-zone model
that helped elucidate the physics of delayed mixed bursts.  They
showed that the competition between nuclear heating due to
triple-$\alpha$ reactions and hot CNO cycle H burning on the
one hand, and radiative cooling via photon diffusion and emission on
the other hand, drives an overstability that eventually triggers a
thin-shell thermal instability and hence a delayed mixed burst.  They
asserted that H burning via the temperature-independent hot CNO
cycle, augmented by the extra seed nuclei produced from stable He
burning, significantly lowers the temperature sensitivity of the total
nuclear energy generation rate to such an extent as to suppress the
thin-shell thermal instability for $\dot{M} \gtrsim 0.3
\dot{M}_{\mathrm{Edd}}$.  Therefore, H burning via the hot CNO
cycle is ultimately responsible for the lower critical $\dot{M}$ above
which bursts do not occur in nature.

The above argument perhaps explains why one-zone models fail, since
the models generally focus only on He burning and make large
approximations with respect to H burning.  But why do detailed
time-dependent multi-zone models with large reaction networks also
perform poorly in relation to observations?  These models are much
more sophisticated than the models of NH03 and CN06 and consequently
ought to perform better, whereas in fact the latter models agree much
better with observations.  CN06 hypothesized that the time-dependent
multi-zone burst models may have used too large a rate for the
experimentally poorly-constrained hot CNO cycle breakout reaction
$^{15}$O($\alpha$,$\gamma$)$^{19}$Ne
\citep{W69,WW81,LWFG86,WGS99,FGWD06}.  We test this hypothesis.  We
begin in \S \ref{themodel} with a description of the model, and we
present the results of the model in \S \ref{results}.  We discuss the
results in \S \ref{discussion}, and we conclude in \S
\ref{conclusions}.

\section{The Model}\label{themodel}

We use the general-relativistic global linear stability analysis of
\citet{CN05}, which is an expanded and improved version of the model
of NH03, to determine the stability of nuclear burning on accreting
neutron stars.  We assume that matter accretes spherically onto a
neutron star of gravitational mass $M =1.4 M_{\odot}$ and areal radius
$R = 10.4$ km at a rate $\dot{M}$, where $\dot{M}$ is the rest mass
accreted per unit time as measured by an observer at infinity.  We set
the composition of the accreted matter to be that of the Sun, such
that at the neutron star surface the H mass fraction
$X_{\mathrm{out}} = 0.7$, He mass fraction $Y_{\mathrm{out}} =
0.28$, CNO mass fraction $Z_{\mathrm{CNO, out}} = 0.016$, and heavy
element fraction $Z_{\mathrm{out}} = 0.004$, where $Z$ refers to all
metals other than CNO.  In this section, we describe the modifications
to the theoretical model.

For temperatures $T \gtrsim 8 \times 10^{7}$ K, H burns
predominantly via the hot CNO cycle
$^{12}$C($p$,$\gamma$)$^{13}$N($p$,$\gamma$)$^{14}$O($\beta^{+}
\nu$)$^{14}$N($p$,$\gamma$)$^{15}$O($\beta^{+}
\nu$)$^{15}$N($p$,$\alpha$)$^{12}$C, the rate of which is determined
by the slow $\beta$-decays of $^{14}$O($t_{1/2}=70.6$ s) and
$^{15}$O($t_{1/2}=122$ s) \citep{HF65}.  During the hot CNO cycle,
essentially all of the CNO ions are converted to $^{14}$O and
$^{15}$O.  For the range of accretion rates we consider in this
investigation, the primary breakout reaction from the hot CNO cycle
into the rp-process of \citet{WW81} is
$^{15}$O($\alpha$,$\gamma$)$^{19}$Ne \citep{GWT95,Hetal96,SBCW99}.  If
this breakout reaction occurs, the $^{19}$Ne($t_{1/2}=17.2$ s)
produced in this reaction can $\beta$-decay and return to the hot CNO
cycle via $^{19}$Ne($\beta^{+} \nu$)$^{19}$F($p$,
$\alpha$)$^{16}$O($p$, $\gamma$)$^{17}$F($p$,
$\gamma$)$^{18}$Ne($\beta^{+} \nu$)$^{18}$F($p$, $\alpha$)$^{15}$O,
the effect of which is only to expedite the hot CNO cycle by a factor
$\lesssim 1.6$.  We ignore this effect in our model.  If $^{19}$Ne
captures a proton, however, the resulting $^{20}$Na ion can never
return to the hot CNO cycle \citep{WW81}.  The fate of a $^{19}$Ne ion
thus depends on whether or not it captures a proton before it can
$\beta$-decay.  Therefore, the probability that a $^{19}$Ne ion is
removed from the hot CNO cycle is
\begin{equation}
R = \frac{\rho X \lambda_{p \gamma}(^{19}\mathrm{Ne})}{\rho X
\lambda_{p \gamma}(^{19}\mathrm{Ne}) +
\lambda_{\beta^{+}}(^{19}\mathrm{Ne})},
\end{equation}
where $\rho$ is the density, $X$ is the H mass fraction,
$\lambda_{p \gamma}(^{19}\mathrm{Ne})$ is the
$^{19}$Ne($p$,$\gamma$)$^{20}$Na reaction rate from \citet[][hereafter
CF88]{CF88} with electron screening from \citet{DGC73}, and
$\lambda_{\beta^{+}}(^{19}\mathrm{Ne}) = \ln 2 / t_{1/2}$ is the
$^{19}$Ne $\beta$-decay rate.  If $^{19}$Ne captures a proton, further
proton captures will ensue until the nuclear flow reaches the first
waiting point of the rp-process, which we assume to be $^{24}$Si
\citep{WSC98}.  The $^{15}$O($\alpha$,$\gamma$)$^{19}$Ne reaction is
the slowest reaction of this sequence, and so its rate governs the
total reaction rate of this flow.  Therefore, we follow \citet{WGS99}
and approximate the total nuclear energy generation rate of the hot
CNO cycle breakout reaction sequence by
\begin{equation}
\epsb = 24 \estarb \left(\frac{Y}{4}\right)
\left(\frac{Z_{\mathrm{CNO}}}{15}\right) \rho R \lambda_{\alpha
\gamma}(^{15}\mathrm{O}) \times \frp,
\end{equation} 
where $\estarb = Q_{\mathrm{rp}}/24 m_{\mathrm{p}} = 1.2 \times
10^{18}$ $\mathrm{ergs\,g}^{-1}$, $Y$ is the He mass fraction,
$Z_{\mathrm{CNO}}$ is the mass fraction of CNO elements, and
$\lambda_{\alpha \gamma}(^{15}\mathrm{O})$ is the
$^{15}$O($\alpha$,$\gamma$)$^{19}$Ne reaction rate from CF88,
which is based on the rate derived by \citet{LWFG86}.  For the
temperatures $T \lesssim 6 \times 10^{8}$ K we consider in this work,
the resonant contribution of the 4.03 MeV state in $^{19}$Ne dominates
the $^{15}$O($\alpha$,$\gamma$)$^{19}$Ne reaction rate \citep{LWFG86}.
The reaction rate of this contribution is directly
proportional to the $\alpha$ width $\Gamma_{\alpha}$.  However,
$\Gamma_{\alpha}$ is very difficult to measure experimentally, and so
the true $\lambda_{\alpha \gamma}(^{15}\mathrm{O})$ is highly
uncertain \citep[for recent progress on experimental constraints, see,
e.g.,][]{Tetal05,Ketal06}.  To account for this uncertainty, we
multiply the CF88 rate by a dimensionless free parameter
$\frp$ such that $\frp=1$ corresponds the the usual rate of CF88.  
We omit the electron screening contribution to the
reaction rate since the enhancement factor itself depends on the
relatively unconstrained resonance strength \citep{SvH69,M77}.
Instead, we simply absorb this contribution into $\frp$.  Equations
(6), (8), (9), and (10) of \citet{CN05} thus become
\begin{equation}
e^{-2 \Phi/c^{2}} \frac{{\partial}}{{\partial}\Sigma} \left (\frac{F
	r^{2}}{R^{2}} e^{2 \Phi/c^{2}} \right ) = T \ddt{s} - (\epsh +
	\epshe + \epsb + \epsilon_{\mathrm{N}} -
	\epsilon_{\mathrm{\nu}}),
\end{equation} 
\begin{equation}\label{dxdteqn}
\ddt{X} = - \frac{\epsh}{\estarh} - \frac{5}{24}\frac{\epsb}{\estarb},
\end{equation} 
\begin{equation}
\ddt{Y} = \frac{\epsh}{\estarh} - \frac{\epshe}{\estarhe} -
\frac{4}{24} \frac{\epsb}{\estarb},
\end{equation} 
\begin{equation}
\ddt{Z_{\mathrm{CNO}}} = \frac{\epshe}{\estarhe} - \frac{15}{24}
\frac{\epsb}{\estarb}.
\end{equation}
See \citet{CN05} for the definitions of the various symbols.  Note
that the factor of $5$ in equation (\ref{dxdteqn}) accounts for the
five proton captures needed to reach $^{24}$Si.  For a substantial
portion of the accretion rates we consider in this work, the nuclear
flow due to stable burning extends well beyond the relatively small
number of isotopes included in our limited network
\citep[e.g.,][]{SBCW99,FBLTW05}.  However, the contributions from
H and He burning dominate the total nuclear energy
generation rate and set the thermal profile of the accreted layer, and
we calculate these contributions to high accuracy.  Consequently, the
thermal profiles resulting from our network compare very well with
those of \citet{SBCW99} for the range of accretion rates in which our
calculations overlap.  Furthermore, we are interested only in the
physics of the burst onset, for which only hot CNO cycle,
triple-$\alpha$, and breakout reactions are important.  Thus, our
limited network should be adequate for our purposes.

\section{Results}\label{results}

\begin{figure}
\epsscale{1.25}
\plotone{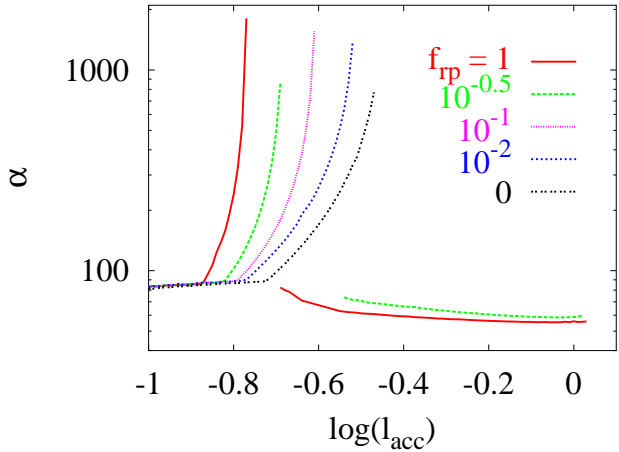}
\caption{Plot shows $\alpha$-values of type I X-ray bursts as a
function of the Eddington-scaled accretion rate $\lacc \equiv
\dot{M}/\dot{M}_{\mathrm{Edd}}$ for five different values of $\frp$,
the dimensionless factor by which we multiply the
$^{15}$O($\alpha$,$\gamma$)$^{19}$Ne reaction rate from CF88.
}
\label{alphagraph}
\end{figure}

Figure \ref{alphagraph} shows the values of the dimensionless quantity
$\alpha$ as a function of the Eddington-scaled accretion rate $\lacc
\equiv \dot{M} / \dot{M}_{\mathrm{Edd}}$ for five different values of
$\frp$, where $\alpha$ is defined as the accretion energy released
between successive bursts divided by the nuclear energy released
during a burst.  We follow NH03 and call a type I X-ray burst
``prompt'' if $\alpha \lesssim 100$ and ``delayed'' if $\alpha \gg
100$.  The case $\frp = 0$ omits the breakout reaction sequence
entirely, and so this calculation is representative of the results of
NH03 and CN06.  The models gives prompt bursts for $\lacc \lesssim
0.15$ and delayed bursts over the range $\lacc \approx 0.15$-$0.3$.
For low values of $\frp \lesssim 0.1$, the effects of breakout
reactions from the hot CNO cycle on the occurrence of type I X-ray
bursts are minor, and the models of NH03 and CN06 continue to provide
an adequate description of the nuclear physics that precedes bursts.
However, the situation changes for $\frp \gtrsim 0.1$. The critical
$\lacc$ above which delayed mixed bursts cease now significantly
decreases, and the range of accretion rates in which delayed mixed
bursts occur becomes notably truncated as well.  More importantly, a
new burst regime appears at accretion rates close to
$\dot{M}_{\mathrm{Edd}}$, where the bursts have $\alpha < 100$ and are
hence prompt.  This regime of bursts near the Eddington limit was
hypothesized by CN06, but it was not present in either the NH03 or
CN06 models due to the exclusion of hot CNO cycle breakout reactions.
The usual delayed mixed burst regime and this new prompt mixed burst
regime are separated by a short range of accretion rates in which
bursts do not occur.  These results obtained with larger values of
$\frp$ agree rather well with previous theoretical models
\citep{FHM81,AJ82,P83,T85,TWL96,B98,FHLT03,HCW05}, most of which
predicted bursts to occur up to roughly the Eddington limit.  Figure
\ref{alphagraph} thus explains the origin of the differences between
those models and the models of NH03 and CN06.  If the breakout
reaction rate is large, say $\frp \gtrsim 0.1$, the results agree with
most published models (which include breakout reactions in full
strength), and if the rate is small, $\frp \lesssim 0.1$, the results
are similar to those obtained by NH03 and CN06 (who effectively set
$\frp = 0$).  We reiterate that the latter results agree much better
with observations.

\section{Discussion}\label{discussion}

For $\lacc \lesssim 0.1$, the temperatures reached during steady-state
nuclear burning in the accreted layer are too low for significant
leakage out of the hot CNO cycle via breakout reactions, and so these
reactions have a negligible effect on the onset of type I X-ray
bursts, regardless of the true $^{15}$O($\alpha$,$\gamma$)$^{19}$Ne
cross section.  However, breakout reactions should affect the nuclear
flow during the burst itself, and hence they could influence the burst
lightcurve and possibly also the onset of subsequent bursts
\citep{FGWD06}.  CN06 showed that the delayed mixed bursts of NH03
occur when $\sh$, the column depth at which H is depleted via
stable nuclear burning as measured from the stellar surface, is less
than but close to $\she$, the column depth at which He is depleted
via stable nuclear burning.  The hot CNO cycle breakout reaction
sequence (i) eliminates seed nuclei from the hot CNO cycle which slows
H burning, thereby increasing $\sh$, and (ii) provides an
additional pathway by which He may burn, thereby decreasing
$\she$.  Consequently, one expects the regime of delayed mixed bursts
to occur at lower $\lacc$ if breakout reactions are included.
Furthermore, CN06 showed that the interplay between H burning
via the hot CNO cycle and He burning via triple-$\alpha$ reactions
is integral to generating the oscillations that precede delayed mixed
bursts.  Breakout reactions diminish this interplay by eliminating the
hot CNO cycle seed nuclei and should therefore reduce the range of accretion
rates over which delayed mixed bursts occur.  Figure \ref{alphagraph}
illustrates both of these effects (compare $\frp=1$ with $\frp=0$).

For accretion rates above those at which delayed mixed bursts occur,
NH03 found that nuclear burning is always stable and therefore bursts
do not occur, in accord with observations
\citep{vPCLJ79,vPPL88,Cetal03,RLCN06}.  According to their model, for
$0.3 \lesssim \lacc \lesssim 1$, steady-state H burning via the
hot CNO cycle increases the effective radiative cooling rate and
thereby suppresses a He-triggered thin-shell thermal instability
(CN06).  However, their model did not include breakout reactions.  If
the $^{15}$O($\alpha$,$\gamma$)$^{19}$Ne breakout reaction rate is
significant in this regime, this reaction will suppress hot CNO cycle
H burning, and thus He burning via triple-$\alpha$
reactions will govern the total reaction rate, since the
triple-$\alpha$ reaction rate will be the slowest rate in the nuclear
flow at these high $\lacc$.  The notion that the triple-$\alpha$
reaction rate is the slowest rate in the nuclear flow is precisely
what one-zone type I X-ray burst ignition models assume
\citep{FHM81,P83,B98,CB00,HCW05}, and we suggest that this is the
reason why the results of one-zone and time-dependent multi-zone
models agree so well.  All of these models predict that nuclear
burning is thermally stable for temperatures $T \gtrsim 5 \times
10^{8}$ K, which are reached only at accretion rates $\lacc \gtrsim 1$
\citep[e.g.,][]{SBCW99}.  Figure \ref{alphagraph} illustrates that,
for relatively large values of $\frp$, there is a regime of prompt
bursts at accretion rates greater than the accretion rates at which
delayed mixed bursts occur, and the critical accretion rate above
which these prompt mixed bursts cease is roughly
$\dot{M}_{\mathrm{Edd}}$, in very good agreement with these other
models.  Figure \ref{alphagraph} illustrates also that nuclear burning
is stable for the small range of $\lacc$ between these two regimes.
These accretion rates are high enough to suppress delayed mixed
bursts, but they are not high enough to cause a sufficient leakage out
of hot CNO cycle H burning and trigger a thermal instability.

The model we present in this work suggests that, if the true
$^{15}$O($\alpha$,$\gamma$)$^{19}$Ne cross section is greater than
approximately $0.1$ of the CF88 rate, type I X-ray bursts should occur
in systems with accretion rates near the Eddington limit.
Observations indicate that this is not the case, since low-mass X-ray
binaries with $\lacc \gtrsim 0.3$ generally do not exhibit bursts.
This suggests that the true $^{15}$O($\alpha$,$\gamma$)$^{19}$Ne
reaction rate is less than that proposed by CF88.  This conclusion is
complementary to that of \citet{FGWD06}, who were the first to propose
that the occurrence of type I X-ray bursts is sensitive to the
strength of the $^{15}$O($\alpha$,$\gamma$)$^{19}$Ne reaction rate,
and who found that the existence of bursts in systems with $\lacc
\approx 0.1$ suggests a lower bound on this rate.  This lower bound
corresponds to $\frp \approx 0.05$ in our notation.  We also note that
a relatively low $^{15}$O($\alpha$,$\gamma$)$^{19}$Ne reaction rate is
consistent with the existence of carbon-triggered superbursts, since a
low rate would increase the carbon yield resulting from stable nuclear
burning \citep{SBCW99,CMSN06,FGWD06}.  The systems GX 17+2 and Cyg X-2
are exceptions to this empirical rule, however, for they exhibit
bursts at accretion rates near the Eddington limit
\citep[e.g.,][]{KG84,THKMM84,KHvdKLM02}.  However, it is possible that
they show bursts for other reasons, e.g., they harbor mass donor stars
with H-deficient envelopes \citep{CMSN06}.

Although the resonant contribution of the 4.03 MeV state dominates the
$^{15}$O($\alpha$,$\gamma$)$^{19}$Ne reaction rate for the
temperatures $T \lesssim 5 \times 10^{8}$ K at which helium burning
can possibly trigger a type I X-ray burst, other resonances such as
the 4.38 MeV state contribute as well.  However, these resonances
contribute much less than a tenth of the 4.03 MeV contribution
\citep[e.g.,][]{Detal03}, whereas a reduction of the total
$^{15}$O($\alpha$,$\gamma$)$^{19}$Ne reaction rate by a factor of only
$\approx 10$ is needed to stabilize nuclear burning.  Therefore, it is
the 4.03 MeV $\alpha$ width alone that determines the stability of
nuclear burning at high accretion rates.

\section{Conclusions}\label{conclusions}

Using the global linear stability analysis of \citet{CN05}, which is
an expanded and improved version of the model of NH03, we have
investigated the effects of the hot CNO cycle breakout reaction
$^{15}$O($\alpha$,$\gamma$)$^{19}$Ne on the occurrence of type I X-ray
bursts on accreting neutron stars at high accretion rates.  For low
values of $\frp \lesssim 0.1$, where $\frp$ is a dimensionless factor
by which we multiply the $^{15}$O($\alpha$,$\gamma$)$^{19}$Ne reaction
rate of CF88, the hot CNO cycle breakout reaction slightly
lowers the critical accretion rate above which delayed mixed bursts
occur, but otherwise the breakout reaction has little effect on the
burst onset.  The predictions of these models are in good agreement
with observations.  For $\frp \gtrsim 0.1$, a new regime of prompt
mixed bursts appears at accretion rates above the rates at which
delayed mixed bursts occur, and the bursts survive up to roughly the
Eddington limit.  The existence of this prompt mixed burst regime up
to the Eddington limit is consistent with nearly all previous
theoretical models.  Our results support the hypothesis of CN06 that
the discrepancies between the results of the global linear stability
analysis of NH03 and the results of time-dependent multi-zone models
with large reaction networks may be caused by the latter models
assuming too large a strength for the
$^{15}$O($\alpha$,$\gamma$)$^{19}$Ne reaction rate at temperatures $T
\lesssim 6 \times 10^{8}$ K.  The fact that observations agree much
better with the results of NH03 and CN06 implies that the true
reaction rate is lower than the rate assumed in these multi-zone
models.  Specifically, we suggest that the true $\alpha$ width
$\Gamma_{\alpha}$ of the 4.03 MeV state in $^{19}$Ne is lower than the
$\alpha$ width proposed by \citet{LWFG86}.  Calculations using
multi-zone models such as those of \citet{Wetal04} and \citet{FGWD06},
but with the $^{15}$O($\alpha$,$\gamma$)$^{19}$Ne reaction rate
lowered in strength, need to be carried out to either confirm or
refute this suggestion.

\acknowledgments

We thank the referee for several insightful comments and suggestions.
This work was supported by NASA grant NNG04GL38G.

\clearpage

\end{document}